\documentclass{article}
\usepackage{spconf,graphicx}
\usepackage{amsmath, amsfonts}
\usepackage{mathrsfs}
\usepackage{bm, braket}
\usepackage{bbm}
\usepackage{float}
\usepackage{xspace}
\usepackage{color}

\def\fig{Fig.}
\def\eqn{Eq.}

\def\tab{Tab.}

\def\eg{{\textit{e.g.}\xspace}}

\def\etc{{\textit{etc}\xspace}}

\graphicspath{{./figures/}}
\graphicspath{{./pdf/}}
\DeclareGraphicsExtensions{.pdf,.jpeg,.png}

\ifodd 1
\newcommand{\com}[1]{\textbf{\color{red}(COMMENT: #1)}} 
\newcommand{\resp}[1]{\textbf{\color{cyan}(Re: #1)}} 
\newcommand{\todo}[1]{\textbf{{\color{orange}(TODO: #1)}}}
\newcommand{\unused}[1]{{\color{gray}#1}}
\else

\newcommand{\com}[1]{}
\newcommand{\resp}[1]{}
\newcommand{\todo}[1]{}
\newcommand{\unused}[1]{}
\fi

\title{Quantum Ranging Enhanced TDoA Localization}
\name{Entong He, Yuxiang Yang, Chenshu Wu}
\address{Department of Computer Science, The University of Hong Kong}
\begin{document}
%
\maketitle
\begin{abstract}
Localization is critical to numerous applications. 
The performance of classical localization protocols is limited by the specific form of distance information and suffer from considerable ranging errors. This paper foresees a new opportunity by utilizing the exceptional property of entangled quantum states to measure a linear combination of target-anchor distances. Specifically, we consider localization with quantum-based TDoA measurements. Classical TDoA ranging takes  the difference of two separate measurements.
Instead, quantum ranging allows TDoA estimation within a single measurement, thereby reducing the ranging errors. Numerical simulations demonstrate that the new quantum-based localization significantly outperforms conventional algorithms based on classical ranging, with over 50\% gains on average. 

\end{abstract}
%


\section{Introduction}
\label{sec:intro}
Numerous techniques have been proposed to achieve accurate localization \cite{wu2018wireless}, including range-based methods and range-free methods. Under the classical range-based localization framework, sensors determine their locations via nodes with known positions, generally referred to as \textit{anchors} \cite{sensorLocMethods}. Existing localization protocols rely on ranging techniques including Time of Arrival (ToA) \cite{TOALoc}, Time Difference of Arrival (TDoA) \cite{TDoALoc}, and Received Signal Strength (RSS) \cite{RSSLoc}, among which the information on distances between the target location and the anchor is encoded into the signal in various ways. Various modalities have been exploited, such as radio signals (\eg, Wi-Fi, RFID, UWB, \etc), acoustic signals, visible light, magnetic signals, \etc. 

Due to the limit of signal transmission form under the classical framework, conventional ranging merely enables measuring single sensor-anchor distances (ToA, RSS, \etc) or the difference of two such distances (TDoA), which corresponds to a circle and a hyperbolic curve on a plane, respectively. In this paper, we provide the first glance to introduce the theory of quantum mechanics \cite{quanMech} and quantum metrology \cite{quanMetro, AdvQuanMetro, BeatQuantumLimit} into the field of localization, which we call quantum-enhanced ranging. Designing a simple entangled probing quantum state allows a \textit{concurrent} distance measurement to multiple anchors, in the form of any $\pm 1$ combination of individual sensor-anchor distances, within one-shot ranging. 
This allows a new localization paradigm with reduced ranging errors.

Localization using quantum ranging, however, raises a non-convex objective minimization problem. We solve the problem by relaxing it into a convex conic optimization \cite{cvxOpt}. Simulations are conducted by simulating erroneous ranging in the TDoA under the quantum ranging framework, and the comparison to classical TDoA and CRLB validates the advantage of our proposed localization technique. The numerical results exhibit that with the aid of quantum ranging, our method significantly outperforms conventional ranging methods, breaking the limit of classical systems.

The rest of the paper is organized as follows: 
\S\ref{sec:sysmodel} introduces quantum ranging, presents the proposed algorithm, and derives a closed-form expression for CRLB. \S\ref{sec:eval} presents the evaluation results, and \S\ref{sec:conclusion} concludes the paper.

\section{System Model}
\label{sec:sysmodel}
\subsection{Quantum Ranging}
\label{subsec:preQER}
We propose a new ranging protocol based on the specific property of the entangled quantum state that incorporates multiple sensor-anchor distances into the total distance within one-shot ranging. The process of obtaining distance information is enabled by the time-variation property of an isolated quantum state. To see this, we first consider the Schr\"odinger equation \cite{quanMech} on a quantum state $\ket{\psi(t)}$ that depicts the evolution of entangled photons state with invariant Hamiltonian $H$ when emitted into the space
\begin{equation}
    \label{eqn:schEq}
    i \hbar \frac{\partial}{\partial t} \ket{\psi(t)} = H \ket{\psi(t)}
\end{equation}
Suppose the initial state $\ket{\psi(0)}$ is prepared to be an eigenstate of $H$ corresponds to eigenvalue $E$ under energy expression, \eqn\eqref{eqn:schEq} is solved by 
\begin{equation}
    \label{eqn:solSchEq}
    \ket{\psi(t)} = \exp\left(-i\frac{E}{\hbar}t \right) \ket{\psi(0)}
\end{equation}
When employing a two-qubit addition probe state $\ket{\psi(0)} = \frac{1}{\sqrt{2}} \left(  \ket{0} + \ket{1} \right)$, with state energy $E_0, E_1$ at states $\ket{0}, \ket{1}$ respectively, the evolved state \cite{quanEvo} can be written as
\begin{equation}
    \label{eqn:evolvedState}
    \ket{\psi(t)} = \frac{1}{\sqrt{2}} e^{-i\frac{E_0}{\hbar} t} \left( \ket{0} + e^{-i\theta} \ket{1} \right):~\theta = \frac{E_1 - E_0}{\hbar} t
\end{equation}
Utilizing the methodologies in quantum phase estimation, the information on the photon's time of flight (TOF) $t$ can be extracted. Typically, a binary quantum measurement scenario $\{\Pi_0, \Pi_1\}$ is utilized \cite{quanPhaseEst, TQI}, where $\Pi_0 = \ket{\psi(0)} \bra{\psi(0)}$ is the projection operator onto the space of the initial probe state, and $\Pi_1 = \textbf{id} - \Pi_0$ where $\textbf{id}$ is the identity operator on the Hilbert space $\mathscr{H}$. The probabilities of obtaining two resulting states are given by
\begin{equation}
\label{eqn:prob}
p_0 = \left< \psi(t)|\Pi_0|\psi(t)  \right> = \cos^2\frac{\theta}{2},~~ p_1 = 1 - p_0 = \sin^2 \frac{\theta}{2}
\end{equation}
Suppose in the $d$-dimensional space, the photon travels from the sensor location $\bm{x}$ and gets reflected by the anchor $\bm{a}$ where $\bm{x}, \bm{a} \in \mathbb{R}^d$. Once get recollected by the sensor, its state is parametrized by the total phase shift $\theta$, from which the total distance can be obtained with the additional light speed $c$.
\par To generalize, we can realize measuring $\pm 1$-combination of the aforementioned sensor-anchor distances. More specifically, suppose a total of $n$ anchors $\{\bm{a}_1, \dots, \bm{a}_n\}$ are available, for the $k^{\text{th}}$ ranging, an index set $I_k \subseteq \{1, \dots, n\}$ picks several anchors from all available ones. For anchor $i \in I_k$, we use $\omega_{i, k} \in \{-1, 1\}$ to indicate the sign assignment. To realize such a ranging scenario, we prepare an entangled probe state correspondingly, which is explicitly expressed as
\begin{equation}
    \label{eqn:probeState}
    \ket{\psi_{\text{init}}} = \frac{1}{\sqrt{2}} \sum_{j = 0, 1} \bigotimes_{i \in I_k} \ket{\pi_{i, k}^j},~~\pi^{j}_{i, k} = \omega_{i, k} + j~\text{mod}~2
\end{equation}
After the independent time-variation of each photon, we arrive at the resulting state
\begin{equation}
    \label{eqn:resultState}
    \begin{aligned}
    \ket{\psi_{\text{ret}}} &= \frac{1}{\sqrt{2}} e^{-i\gamma}  \left( \bigotimes_{i \in I_k} \ket{\pi_{i, k}^0} + e^{-i\theta} \bigotimes_{i \in I_k} \ket{\pi_{i, k}^1} \right) \\
    \gamma &= \frac{E_0}{\hbar} \sum_{\substack{i \in I_k \\ \omega_{i, k} = 1}   } \frac{2\|\bm{x} - \bm{a}_i\|}{c} + \frac{E_1}{\hbar} \sum_{\substack{i \in I_k \\ \omega_{i, k} = -1}} \frac{2\|\bm{x} - \bm{a}_i\|}{c} \\ \theta &= \frac{2(E_1 - E_0)}{\hbar c} \left( \sum_{i \in I_k} \omega_{i, k} \|\bm{x} - \bm{a}_i\| \right)
    \end{aligned}
\end{equation}
By applying an analogous binary measurement scenario, the estimation on $\theta$ can be done based on the phase-probability relation as shown in \eqn\eqref{eqn:prob}. The eventual distance combination $\sum_{i \in I_k} \omega_{i, k} \|\bm{x} - \bm{a}_i\|$ can thus be obtained by computing $\hbar c \theta / 2(E_1 - E_0)$.

\subsection{Localization through Relaxation}
\label{subsec:locAlgo}
The geometry of our proposed quantum ranging localization can be stated as follows: On a $\mathbb{R}^d$ plane there are a set $A$ of $n$ anchors $A = \{\bm{a}_1, \dots, \bm{a}_n\}$ are available, a sensor with unknown location $\bm{x}$ attempts to determine its position through $m$ times of quantum ranging with results $\bm{d} = (d_1, \dots, d_m)$. In the presence of phase estimation error, each distance can be modeled as $d_k = \hat{d_k} + \epsilon_k$ where $\epsilon_k: 1 \leq k \leq m$ are independent and identically distributed (i.i.d.) zero-mean Gaussian noise. Under these assumptions, the localization problem can be formulated as the following optimization problem:
\begin{equation}
    \label{eqn:noncvx}
    \min_{\bm{x}} \sum_{k=1}^m \left(  \sum_{i \in I_k} \omega_{i,k} \|\bm{x} - \bm{a}_i\| - d_k \right)^2
\end{equation}
To realize the TDoA ranging, we restrict $|I_k| = 2$ for all $1 \leq k \leq m$, and the two anchors included by $I_k$ are assigned to different signs. Let $I_k = \{ \alpha_{k, 1}, \alpha_{k, 2} \}$ where $\alpha_{k, v}, v = 1, 2$ indicates the anchor index included. The \eqn \ref{eqn:noncvx} is non-convex, but can be transformed into a conic convex optimization problem. For simplicity, auxiliary variable $\bm{y} \in \mathbb{R}^n$ is introduced to linearize the objective function. Let incidence matrix $P \in \mathbb{R}^{m \times n}$, the problem can be written as follows:
\begin{equation}
    \label{eqn:primitveProb}
    \begin{array}{cc}
        \min\limits_{\bm{x}, \bm{y}} &  \|P \bm{y} - \bm{d}\|^2 \\
         \text{s.t.} & y_i = \|\bm{x} - \bm{a}_i\| \\
         &P = \begin{pmatrix}
             \bm{e}(\alpha_{1, 1}, \alpha_{1, 2}) & \cdots & \bm{e}(\alpha_{m, 1}, \alpha_{m, 2})
         \end{pmatrix}^T \\
         &\forall ~~ 1 \leq i \leq n, \quad 1 \leq k \leq m
    \end{array}
\end{equation}
where the notation $\bm{e}(i, j)$ stands for the vector in $\mathbb{R}^n$ that takes value $1$ at its $i^{\text{th}}$ entry, value $-1$ at its $j^{\text{th}}$ entry, and remaining entries being $0$.
In order to deal with the non-convex term in the constraint, $\bm{y}$ can be raised to a higher dimensional space by introducing $\bm{Y} \in \mathbb{S}^{n}$ and $\gamma \in \mathbb{R}^+$. By combining the relaxation techniques proposed independently by Tseng \cite{SOCPLoc} and Biswas \textit{et al.} \cite{SDPLoc} in the sensor network localization problem, we can reformulate \eqn\ref{eqn:primitveProb} as follows:
\begin{equation}
    \label{eqn:cvxLocProb}
    \begin{array}{cc}
        \min\limits_{\bm{x}, \bm{y}, \bm{Y}, \gamma} & P^TP \bullet \bm{Y} - 2\bm{d}^T P \bm{y} \\
         \text{s.t.} & Y_{i,i} = \gamma - 2\bm{a}_i^T \bm{x} + \|\bm{a}_i\|^2, \quad \forall 1 \leq i \leq n \\
         &Y_{i, j} \geq \left| \gamma - (\bm{a}_i + \bm{a}_j)^T \bm{x} + \bm{a}_i^T \bm{a}_j \right|, ~ \forall 1 \leq i < j \leq n \\
         & \gamma \geq \|\bm{x}\|^2, \quad y_i \geq \|\bm{x} - \bm{a}_i\|, \quad \forall 1 \leq i \leq n \\
         &\begin{bmatrix} \bm{Y} & \bm{y} \\ \bm{y}^T & 1 \end{bmatrix} \succeq 0, \quad \bm{Y} \in \mathbb{S}^n
    \end{array}
\end{equation}
where the operation $A \bullet B$ stands for $\text{Tr}(A^TB)$ for symmetric matrices $A, B$, and $\mathbb{S}^n$ is the space of $n \times n$ real symmetric matrices. \eqn\eqref{eqn:cvxLocProb} is a fusion of semidefinite programming (SDP) and second-order cone programming (SOCP), and can be equivalently stated as a pure SDP problem. It contains decision variables and constraints both up to scale $O(n^2)$. An interior-point method \cite{IntPtSDP} can be applied to solve it efficiently.

\subsection{Cram\'er-Rao Lower Bound on Localization Error}
It is known that the noise is distance dependent, thus it is convenient to assume that $\epsilon_k \propto {d}_k$. We suppose the scalar $\eta := \epsilon_k / {d}_k \in [0, 1]$ is set to be identical for all measurements $k \in \{1, \dots, m\}$. By the assumption that $\epsilon_k \sim \mathcal{N}(0, \eta^2 {d}_k^2)$ as previously suggested, the probability density function (PDF) of obtaining distance $\bm{d}$ conditioned on real location $\bm{x}$ can be expressed as:
\begin{equation}
    \label{eqn:pdf}
    \begin{array}{c}
        p(\bm{d}|\bm{x}) = \\
\prod\limits_{k=1}^m \frac{1}{\sqrt{2\pi} \eta {d}_k} \exp\left\{- \frac{ (\|\bm{x} - \bm{a}_{\alpha_{k, 1}}\| - \|\bm{x} - \bm{a}_{\alpha_{k, 2}}\| - d_k)^2}{2 \eta^2 {d}_k^2} \right\}
        \end{array}
\end{equation}
An estimator $\hat{\bm{x}}$ on $\bm{x}$ can be obtained by applying the maximum likelihood estimation (MLE) method, \textit{ie}, $\hat{\bm{x}} := \mathop{\arg\max}_{\bm{x}} p(\bm{d}|\bm{x})$. Maximizing the likelihood function is equivalent to solving \eqn\ref{eqn:noncvx} with a weighted version of the objective function, \textit{i.e.}, the objective function is written as $P^T W^2 P \bullet \bm{Y} - 2 \bm{d}^T W P \bm{y}$ where $W = \text{diag}\left\{ 1/d_1, \dots, 1 / d_m  \right\}$. 
\par In the relaxed problem, all parameters are linearized, and thus the estimator $\hat{\bm{x}}$ induced is an unbiased one. The Cram\'er-Rao Lower Bound (CRLB) allows us to derive a lower bound for the localization technique performance. The Fisher Information Matrix (FIM) $J \in \mathbb{R}^{d \times d}$ is defined as follows \cite{FoundSigProc}:
\begin{equation}
    \label{eqn:FIM}
    \begin{aligned}
    J &= - \mathbb{E} \left[ \frac{\partial^2 \log p(\bm{d}|\bm{x})}{\partial \bm{x} \partial \bm{x}^T}  \right]  \\
    &= \sum_{k=1}^m \frac{1}{\eta^2d_k^2} \left( \frac{\bm{x} - \bm{a}_{\alpha_{k, 1}}}{\|\bm{x } - \bm{a}_{\alpha_{k, 1}}\|} - \frac{\bm{x} - \bm{a}_{\alpha_{k, 2}}}{\|\bm{x} - \bm{a}_{\alpha_{k, 2}}\|} \right)  \\
    & \times \left( \frac{\bm{x} - \bm{a}_{\alpha_{k, 1}}}{\|\bm{x } - \bm{a}_{\alpha_{k, 1}}\|} - \frac{\bm{x} - \bm{a}_{\alpha_{k, 2}}}{\|\bm{x} - \bm{a}_{\alpha_{k, 2}}\|} \right)^T
    \end{aligned}
\end{equation}
An intuitive metric for localization performance can be given as follows: Suppose in $r$ times of simulations under the same testbed, the real sensor position is $\bm{x}^{(l)}$ while the solution returned by arbitrary algorithm is $\hat{\bm{x}}^{(l)}$. We define the mean error (ME) to specify the solution deviation:
\begin{equation}
    \label{eqn:AEN}
    \text{ME} := \frac{1}{r} \sum_{l=1}^r \left\|\bm{x}^{(l)} - \hat{\bm{x}}^{(l)}\right\|
\end{equation}
A non-trivial lower-bound for the expectation of \text{ME}~in repeated experiments, by Jenson's inequality and the CRLB, can be stated as follows:
\begin{equation}
    \label{eqn:errorLB}
    \begin{aligned}
    \mathbb{E} \left[ \text{ME} \right] &= \frac{1}{r}  \sum_{l=1}^r \mathbb{E} \left[ \left\|\bm{x}^{(l)} - \hat{\bm{x}}^{(l)}\right\| \right] \\
    &
    \overset{\text{Jenson}}{\geq} \frac{1}{r} \sum_{l=1}^r \sqrt{\mathbb{E}\left[ (\bm{x}^{(l)} - \hat{\bm{x}}^{(l)})^T(\bm{x}^{(l)} - \hat{\bm{x}}^{(l)}) \right] }  \\
    &\overset{\text{CRLB}}{\geq} \frac{1}{r} \sum_{l=1}^r \sqrt{\text{Tr}\left( J_l^{-1} \right)}
    \end{aligned}
\end{equation}
Where $J_l$ is the FIM corresponds to the $l^{\text{th}}$ experiment, determined by the ground truth $\bm{x}^{(l)}$ and the noise factor $\eta$. 

\section{Evaluation}
\label{sec:eval}

\subsection{Simulation Settings}
In this section, we perform simulation results based on the aforementioned relaxed problem in the 3-dimensional space. The whole testbed is set on a classical computer, where the impact of quantum phase estimation error is simulated by adjusting the noise coefficient $\eta$ throughout the experiments. Both sensors and anchors are restricted in a cubic region, where for all sensor position $\bm{x}$ and anchor deployment $\bm{a}$, it holds that $\bm{x} \in [-D_x, D_x]^3$ and $\bm{a} \in [-D_a, D_a]^3$ for some fixed scalars $D_x, D_a \in \mathbb{R}^+$. 
To include both the near-field ($\bm{x} \in \text{conv}(\bm{a}_1, \dots, \bm{a}_n)$) and far-field ($\bm{x} \not\in \text{conv}(\bm{a}_1, \dots, \bm{a}_n)$) cases, where $\text{conv}(\cdot)$ stands for the convex hull of a set of points on $\mathbb{R}^3$, we set $D_x > D_a$. At each noise level, $r=2000$ simulations are conducted. 
\par Moreover, as is mentioned in the previous context, we restrict to explore the improvement of quantum-assisted localization under an analogous signing to the TDoA, \textit{i.e.}, distances $\|\bm{x} - \bm{a}_i\| - \|\bm{x} - \bm{a}_j\|$ for $i \neq j$ are measured at each time. Details of parameter values are listed in \tab \ref{table:params}.
\begin{table}[H]
\centering
\caption{Parameters of testbed geometry}
\label{table:params}
\begin{tabular}{|c|c|}
\hline
\textbf{parameter}        & \textbf{value}                     \\ \hline
$D_x$            & $2.0$ (m)                   \\ \hline
$D_a$            & $1.0$ (m)                    \\ \hline
$n$              & $16$                       \\ \hline 
$m$              & $8$                         \\ \hline
$\begin{array}{c} \text{anchor} \\ \text{placement} \end{array}$ & \footnotesize $\left\{   \begin{array}{c} 
(1, 1, 1), (-1, 1, 1), (1, -1, 1) \\
(-1, -1, 1), (1, 1, -1), (-1, 1, -1) \\
(1, -1, -1), (-1, -1, -1), (0, 0, 1) \\
(0, 0, -1), (0, 1, 0), (0, -1, 0) \\
(1, 0, 0), (-1, 0, 0) \\
(0, 0, 0.5), (0, 0, -0.5)

\end{array}   \right\}$ \normalsize \\ \hline
$\begin{array}{c}
     \text{ranging} \\ \text{scenario} \\ (\text{$y_i := \|\bm{x} - \bm{a}_i\|$})
\end{array}$ & \footnotesize $ \left\{   
\begin{array}{c}
    y_1  - y_2, y_3 - y_4, y_5 - y_6 \\
    y_7 - y_8, y_9 - y_{10}, y_{11} - y_{12} \\
    y_{13} - y_{14}, y_{15} - y_{16}
\end{array}
\right\}$  \normalsize \\  \hline
$\eta$           & $0\% - 4\%$, step size $1\%$              \\ \hline
\end{tabular}
\end{table}
In the framework of TDoA with quantum ranging, where the distance error originates merely from the estimation of phase shift, the error is single-fold and proportionate to the total distance $d_{ij} = \|\bm{x} - \bm{a}_i\|-\|\bm{x} - \bm{a}_j\|$. In the classical TDoA ranging, the two distances $\|\bm{x} - \bm{a}_i\|$ and $\|\bm{x} - \bm{a}_j\|$ are separately distributed, and the resulting error of $d_{ij}$ is the superposition of term-wise errors.

By the above assumption, we model the noise integrated into distance ranging in following two ways, which subsequently correspond to two TDoA setups:
\begin{itemize}
    \item \textbf{Quantum-assisted TDoA.} Under noise level $\eta$, the noisy distance can be modelled as $d_k = (\|\bm{x} - \bm{a}_{\alpha_{k, 1}} \| - \|\bm{x} - \bm{a}_{\alpha_{k, 2}}\|)(1 + \eta \varepsilon_k)$, where $\varepsilon_k \sim \mathcal{N}(0, 1)$. 
    \item \textbf{Classical TDoA.} Each sensor-anchor distance is integrated with noise, \textit{i.e.}, the noisy distance is expressed as
    $
    d_k = \|\bm{x} - \bm{a}_{\alpha_{k, 1}}\|(1 + \eta \varepsilon_{k, 1}) - \|\bm{x} - \bm{a}_{\alpha_{k, 2}}\|(1 + \eta \varepsilon_{k, 2})
    $, where $\varepsilon_{k, v} \sim \mathcal{N}(0, 1)$ for all $k$ and $v \in \{1, 2\}$. It will serve as a baseline for the performance of our proposed quantum-ranging localization.
\end{itemize}

It is worth noticing that since the difference between two distances is measured, for any $\bm{y}^*$ falling into the feasible region, $\bm{y} := \bm{y}^* + \tau \bm{1}$ will yield another feasible $\bm{y}$ with respect to the objective function, where $\bm{1}$ is the vector with all entries being $1$ in the $\mathbb{R}^n$ space. To avoid large deviations of solution returned by the solver from the optimal one, we add a penalty term to the primitive objective function in \eqn\eqref{eqn:cvxLocProb}. With penalty coefficient $\delta \in \mathbb{R}^+$, the problem now is
\begin{equation}
    \label{eqn:penaltyProb}
    \begin{array}{cc}
        \min\limits_{\bm{x}, \bm{y}, \bm{Y}, \gamma} & P^TP \bullet \bm{Y} - 2\bm{d}^T P \bm{y} + \delta \cdot \text{Tr}(\bm{Y}) \\
         \text{s.t.} & Y_{i,i} = \gamma - 2\bm{a}_i^T \bm{x} + \|\bm{a}_i\|^2, \quad \forall 1 \leq i \leq n \\
         &Y_{i, j} \geq \left| \gamma - (\bm{a}_i + \bm{a}_j)^T \bm{x} + \bm{a}_i^T \bm{a}_j \right|, \forall 1 \leq i < j \leq n \\
         & \gamma \geq \|\bm{x}\|^2, \quad y_i \geq \|\bm{x} - \bm{a}_i\|, \quad \forall 1 \leq i \leq n \\
         &\begin{bmatrix} \bm{Y} & \bm{y} \\ \bm{y}^T & 1 \end{bmatrix} \succeq 0, \quad \bm{Y} \in \mathbb{S}^n
    \end{array}
\end{equation}

The CRLB as a function of the ground truth of sensor location and noise factor for each single randomly generated test case is also computed to serve as a benchmark, against which the efficiency of our proposed algorithm is tested. 

\subsection{Numerical Results}
Our code written in Python generates the simulation settings and solves the mixed SOCP/SDP problem and baselines. It calls MOSEK through cvxpy (Version 1.3) when solving the conic programming problems (proposed algorithm and pure SDP), and runs on a 3.2 GHz AMD Ryzen 7 5800H processor and 16 GB RAM. The aforementioned penalty coefficient $\delta$ is set to be $6 \times 10^{-7}$ throughout the experiment. 
\par
\begin{figure}[htbp]
    \centering
    \label{fig:error}
    \includegraphics[width=0.83\linewidth]{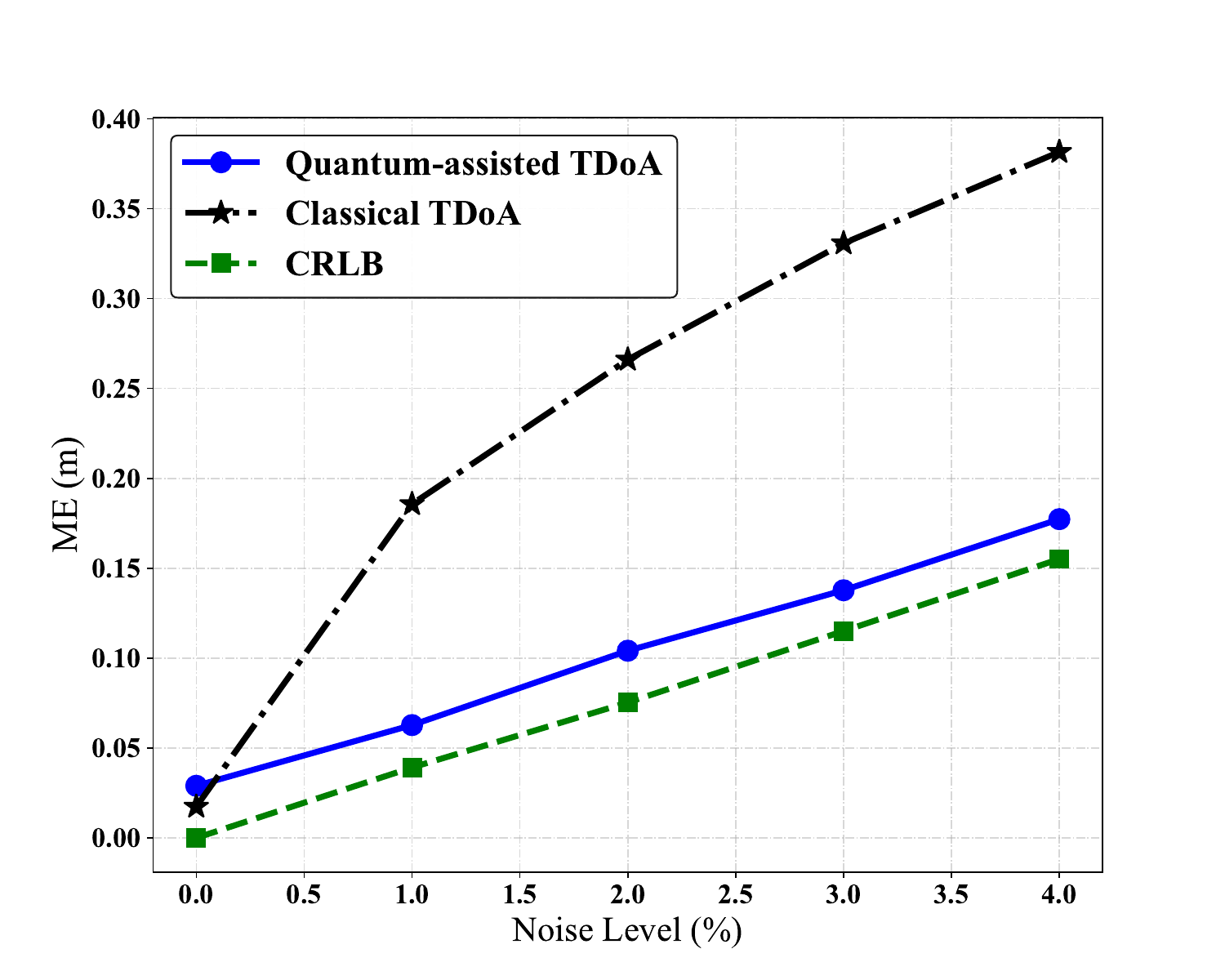}
    \caption{ME under various settings with respect to noise}
    \label{fig:errorNoise}
\end{figure}

 \fig\ref{fig:errorNoise} shows the performance measured by ME of the result solved by our proposed algorithm \eqn\eqref{eqn:cvxLocProb} with penalty under both quantum and classical settings. It can be observed that apart from the noiseless case, \textit{i.e.}, $\eta = 0$, TDoA with quantum ranging consistently outperforms that with traditional ranging. With the increase in noise level, the performance of quantum-assisted TDoA is resilient and the localization error yields a slow trend of increase, which is analogous to the trend of CRLB.

    {The underlying optimization problem yields decision variables and constraints both up to scale $O(n^2)$. To the best of our knowledge \cite{fastSDPAlgo}, a solver would return the optimal solution within $O(n^{5.246})$ iterations. \tab 2 shows the respective runtime of the localization algorithms under different ranging settings.}
It can be observed that our proposed technique trades off computational efficiency for localization accuracy. Rather than the algorithm utilizing SOCP/SDP we suggested, simpler algorithms of higher efficiency can be designed for specific ranging scenarios, which provides a direction for future research.
\begin{table}[H]
    \centering
    \label{tab:timecost}
    \small
    \begin{tabular}{|c|c|}
    \hline
    \textbf{Algorithm}             & \textbf{Average time cost (sec)} \\ \hline
    quantum-assisted ranging    & $0.4348$  \\ \hline
    classical ranging  & $0.4334$ \\ \hline
    \end{tabular}
    \normalsize
    \caption{Time cost of algorithms throughout the simulation}
\end{table}

\section{Conclusion}
\label{sec:conclusion}

In this paper, we describe the first TDoA localization technique under the framework of quantum-enhanced distance ranging. Conic relaxation methods are deployed to solve the raised non-convex optimization problem. The analysis of CRLB is conducted to provide a theoretical lower bound for the localization model. The numerical simulations demonstrate that the proposed algorithm significantly outperforms and breaks the limit of TDoA localization algorithms with classical ranging. Future work shall include designing optimal measurement scenarios when illumination in the environment is taken into account, finding generic ranging scenarios with specific anchor placement geometry, and developing more efficient localization algorithms. 

\section{Acknowledgments}
This work is supported in part by National Natural Science Foundation of China (NSFC) under grant No. 62222216, Guangdong Basic and Applied Basic Research Foundation (Project No. 2022A1515010340) and by the Hong
Kong Research Grant Council (RGC) through the Early
Career Scheme (ECS) grant 27310822, the General Research Fund (GRF) grants No. 17303923.

\vfill\pagebreak


\clearpage
\bibliographystyle{IEEEbib}
\bibliography{refs}

\end{document}